# Design Considerations of a Sub-50 µW Receiver Front-end for Implantable Devices in MedRadio Band

Gregory Chang, Shovan Maity, Baibhab Chatterjee and Shreyas Sen, *Member, IEEE*
School of Electrical and Computer Engineering, Purdue University, West Lafayette, Indiana – 47907, USA
E-mail: {changg, maity, bchatte, shreyas}@purdue.edu

*Abstract*—Emerging health-monitor applications, such as information transmission through multi-channel neural implants, image and video communication from inside the body etc., calls for ultra-low *active* power (<50µW) high data-rate, energy-scalable, highly energy-efficient (pJ/bit) radios. Previous literature has strongly focused on low *average* power duty-cycled radios or low-power but low-date radios. In this paper, we investigate power performance trade-off of each front-end component in a conventional radio including active matching, down-conversion and RF/IF amplification and prioritize them based on highest performance/energy metric. The analysis reveals 50Ω active matching and RF gain is prohibitive for 50µW power-budget. A mixer-first architecture with an N-path mixer and a self-biased inverter based baseband LNA, designed in TSMC 65nm technology show that sub 50µW performance can be achieved up to 10Mbps (< 5pJ/b) with OOK modulation.

## I. INTRODUCTION

IMPLANTABLE devices for healthcare monitoring applications often require to operate at extremely low power. Applications such as multichannel neural implants [1] and ingestible image/video transceivers [2] demand high rate communication. Past works in Medical Device Radio Communication (MedRadio) spectrum around 400 MHz had offered promising opportunities in health care monitoring through extensive use of medical telemetry at low-data-rate [3]. Yet, low-power high data-rate applications continues to be a challenge as transceiver system are size constrained and energy-sparse. For implantable use, to avoid frequent surgeries, batteries must sustain device operation for years; hence, RF transceivers must consume low power during communication. Other requirements such as small form factor, channel selectivity and interference rejection capability still remain crucial [4]. These stringent requirement poses the need to 1re-scrutinize receiver architectural performance trade-offs under the given communication standards and silicon process technology limits. In [5], [6], it is shown that a high data-rate RF receiver typically consumes 1 mW@ 1 Mbps, translating to 1nJ/b energy-efficiency. Hence, improvement in energy-efficiency of low-power high data-rate wireless communication system is the top priority. While recent efforts in pJ/b BAN also include human body communication [7]–[9], we will focus on standard wireless radio in this paper.

In terms of standards, short-range transmission on 401-406 MHz MedRadio band is suitable for the design space of low power operation. Since communication range rarely exceeds 3 m in a BAN, sensitivity requirements are relaxed [10]. The free-space line-of-sight path loss (FSPL) at 403.5 MHz can be calculated using Friis's formula [11], as shown in [12], while keeping the maximum effective isotropic radiated power (EIRP) at -16 dBm for transmitter as per MedRadio standards, the minimum sensitivity required at the receiver is about -64 dBm, which enables lower power operation. Spectrally efficient but energy inefficient modulation schemes are not desirable in such applications. Higher order modulation schemes require higher $E_b/N_0$ and complex decoding, hence higher power. With this constraint, BPSK or OOK is most suitable, while OOK allows simpler and lower power receiver implementation.

The goal of this paper is to explore the performance trade-offs of an Ultra Low Power (ULP) receiver front-end and to prompt for an architecture suitable for sub-50 µW receiver implementation in the TSMC 65nm technology. Here we present 1) the energy-cost trade-off trends of each front-end component such as data bandwidth and sensitivity trades-off with power, 2) justification that mixer-first is a desirable architecture as energy-cost of active matching to 50 ohms is high and 3) with only baseband amplification 3.4pJ/bit energy efficiency is achieved in simulation with OOK transmission.

## II. POWER VS. OPERABLE FREQUENCY TRADE OFF LIMITATION

### A. Low Noise Amplifier (LNA) Design Choice

Conventional LNA often employs inductive source degeneration. This structure is known for improved noise performance and provides passive matching. However, for limited power budget operation, this structure suffers from low transconductance ($g_m$), impacting noise and gain performance. In addition, a larger inductor would be required for proper matching [13]. In many ULP systems, the popular choice of LNA structure is the inverter-based resistive feedback topology [13]-[14]. As shown in Figure 1, signal is also fed to the active load such that bias current is reused but the gain is boosted by the active load's $g_m$. It had also been shown in [13] that due to increased $g_m$, such topology has a better noise figure performance. In [15]-[16], low power consumption was achieved through Ultra-Low-Voltage (ULV) designs at $V_{DD}$=0.4V and 0.5V, respectively. However, the LNA power consumption is about 100µW in both designs.

To evaluate power and operable bandwidth trade-off of the LNA shown in Figure 1, the design was simulated in TSMC 65nm technology. The gate biasing of the NMOS is implemented by the decoupling capacitor $C_{PN}$, separating bias DC levels from both the signal input and LNA output. The mobility ratio of electrons and holes in 65 nm technology is taken into consideration to set the widths of the PMOS and NMOS transistors, so that node $RF_{OUT}$ will be biased at roughly $V_{DD}$ / 2. It is desired to have power consumption of the LNA

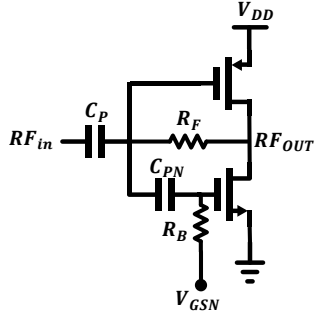

Figure 1: RF/BB Scalable LNA design with NMOS Gate Control

below 30 µW for our target applications.

*1) Energy Cost of Active Matching*

Matching is an important factor in RF system. As shown in [17], active matching can be employed by using a shunt feedback. However, extra power is required for such shunt feedback. For the design choice of Figure 1, matching is realized through the feedback resistor $R_f$. It can be shown analytically that in order to match to 50 Ω, high power is required as it would be limited by technology's $g_m r_o$. The input impedance shown in Eq. 2, which is simplified to Eq. 3.

$$Z_{in} = \frac{R_f}{1+g_m r_o} + \frac{r_o}{1+g_m r_o} \quad (2)$$

$$Z_{in} = \frac{1}{1+k}(R_f + \frac{k}{g_m}) \quad (3)$$

Where $k = g_m r_o$, $g_m = g_{m,n} + g_{m,p}$, and $r_o = r_{o,p} || r_{o,n}$. Given the $k$ value for a particular technology, input impedance matching can be achieved from (3) by choosing appropriate value of $R_f$ and transistor width which controls $k/g_m$ under a given gate bias voltage. In [14] and [18], implementation is done through 180 nm and 130 nm technologies where $g_m r_o$ is higher than to 65 nm. When biasing at optimum $\frac{g_m f_t}{I_d}$, $k$ is around 10~11 for 65 nm. Hence when matching to 50 Ω, it could easily require double the power as compared to 180 nm, as it requires $\left(R_f + \frac{k}{g_m}\right) = 550$. Although value of $R_f$ does not directly impact power, large transistors that supports high current must be used in order to reduce the second term. With this principle, the LNA is shown to perform at various input impedance and gate bias. For low resistance matching such as 50 Ω, the PMOS is easily scaled to 600µm and NMOS at 300µm so that $k/g_m$ term is less than 300. Figure 2 shows the power, bandwidth, $R_f$ required and noise figure achieved at various input impedance. When biasing at higher $V_{GS,N}$, higher bandwidth is achieved from the same $Z_{in}$, but more power will then be consumed. This leads to an interesting observation between power and bandwidth trade off at different $Z_{in}$ matching schemes. In order to achieve low impedance, more power is consumed. Low impedance also exhibits excess bandwidth for a 400MHz band receiver.

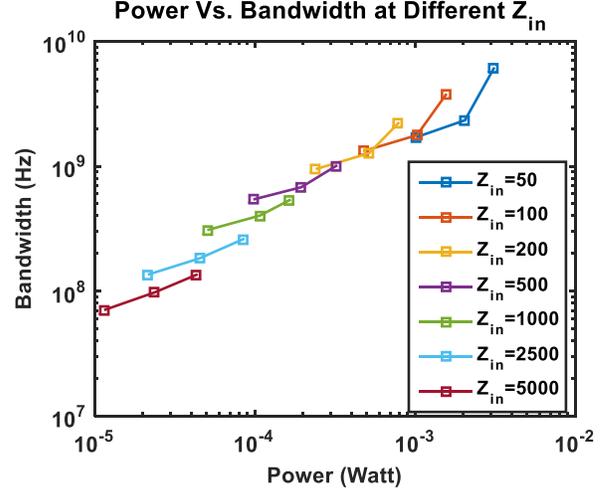

Figure 3: Power vs Bandwidth of LNA at different $Z_{in}$ configuration. High active power is needed for 50 ohms matching and entails excess bandwidth. For µW operations, active matching need to be at kilo-ohms level and utilize passive matching network, avoiding active matching to 50 ohms at RF. The three points along each segment is derived from the three biasing condition $V_{gs}$ shown in Figure 2.

Performance tunability is largely achieved by controlling the gate bias. Derived from Figure 2, Figure 3 shows that matching to 50 Ω easily requires at least 1 mW power and will obtain GHz bandwidth. For a MedRadio transceiver design, such LNA would prove to be overly power-hungry. Further reduction in overhead bandwidth can be achieved through deep sub-threshold gate bias below 0.45V, but this would require unpractically large transistor device and can exceed technology allowed maximum for 65nm, hence only three gate biasing choices were investigated in Figure 2 and Figure 3.

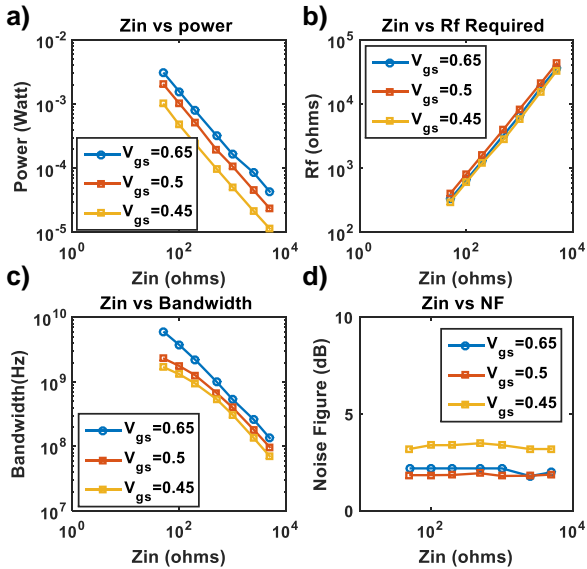

Figure 2: a) Power required for a given $Z_{in}$, showing that at different LNA gate bias, increasing power was required for lower impedance matching. b) Feedback resistor varies proportionally with given $Z_{in}$ as accordance to equation (2). c) Bandwidth increases as $Z_{in}$ is lowered, leads to an interesting observation between power and bandwidth trade off at different $Z_{in}$. d) Noise figure of the LNA under any $Z_{in}$ in general does not vary under matched condition.

*2) Energy Cost of BW*

The power consumption for the LNA varies directly to its bandwidth, as most of the power consumption results from the quiescent current used to bias the circuit. A higher quiescent current alleviates the short-channel resistance $r_o$, lowering any RC combination at output node for voltage-voltage mode

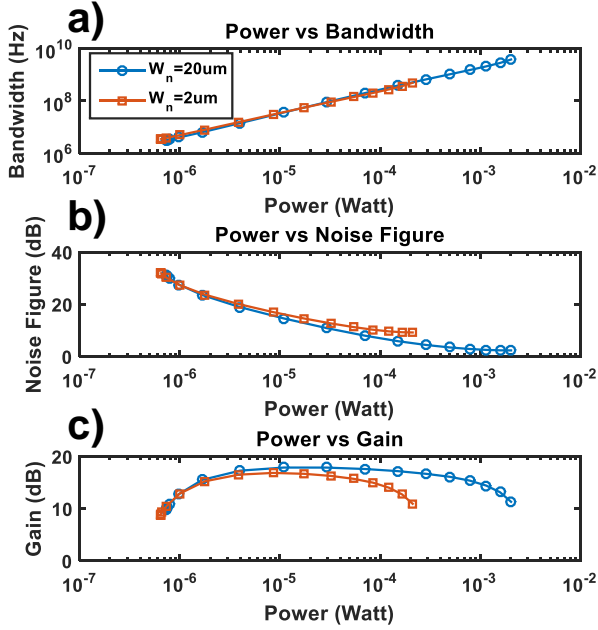

Figure 4: a) LNA's power varies proportionally with bandwidth, where large size LNA can achieve higher bandwidth range. b) Noise figure improves as more power is consumed. c) The flat-band gain increases initially and then saturates with increasing power consumption

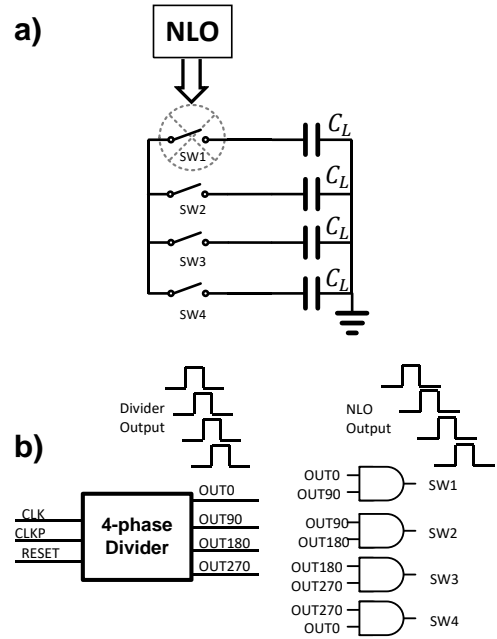

Figure 5: a) 4-Path switches perform passive downconversion controlled by NLO, switch is implemented with NMOS. b) NLO phase generation circuit. Divider generates 90 degree apart 4-phase signl, then fed to AND gates to convert into non-overlapping 4-phase signals.

operation, expanding the bandwidth closer to the technology allowed $f_t$.

As active matching had proved to be power hungry, it motivates LNA for baseband usage in radio front-end rather than RF usage. For BB-LNA, power and bandwidth trade-off trend still exits. As shown in Figure 4 power adjustment can be done through sizing and gate voltage tuning. If bandwidth is limited to MHz range, sub-50 μW implementation can be achieved. This illustrates an important bandwidth-power trade off and shows that it is the desirable design for baseband LNAs.

### B. Frequency Down conversion with N-path Passive Mixer

Frequency translation are another crucial component in power budgeting a receiver. As described in [19], N-path filtering with passive mixers were first proposed as RF frequency bandpass filter. This technique is now often used for narrow band down-conversion by utilizing the frequency translation capability. For such down-conversion design, N parallel paths of passive switches are connected to one input and each path is loaded with a capacitor. To down-convert properly, a non-overlapping LO (NLO) is used to control the mixing operation, such that the baseband capacitor is connected to only one path at a time. The power consumed to operate such device varies with both number of path used, size of switches and operation frequency.

In this implementation, NLO is produced from a divider circuit functioning as the phase generator from a LO source. For front-end power budgeting, LO generation is not considered but phase generation is. For this purpose, two divider architectures shown in Figure 6 were investigated : a conventional flip-flop based divider and a circular divider presented in [20]. Both divider generates 4-phase LO that are 90 degrees apart. These signals were fed to AND gates to produce 25% duty cycle non-overlapping LO as shown in Figure 5b. Figure 7 shows the power consumption comparison between the circular and flip-flop based divider architectures. It can be seen that at frequency of interest for RF receiver the circular divider consumes less power and hence is used for designing this receiver.

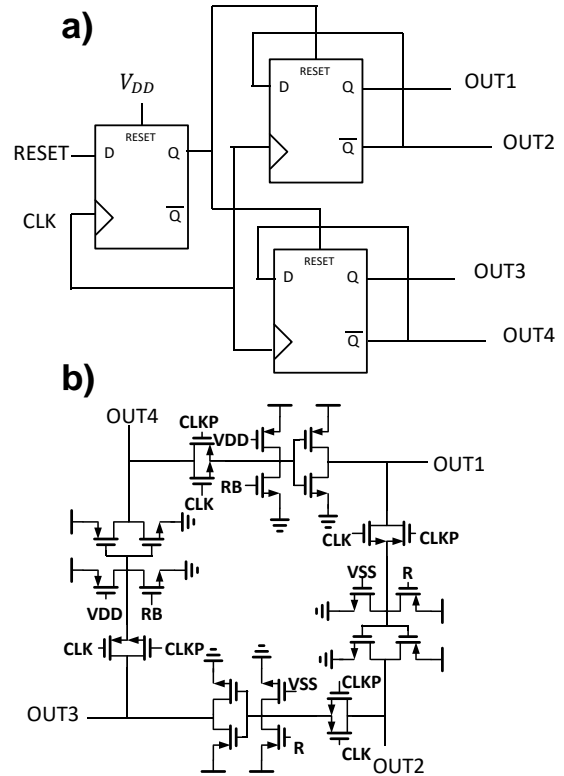

Figure 6: a) Flip-flop based divider. b) Circular divider of [16]. Out1, out2, out3 and out4 are 4-phase 90 degree offset signal

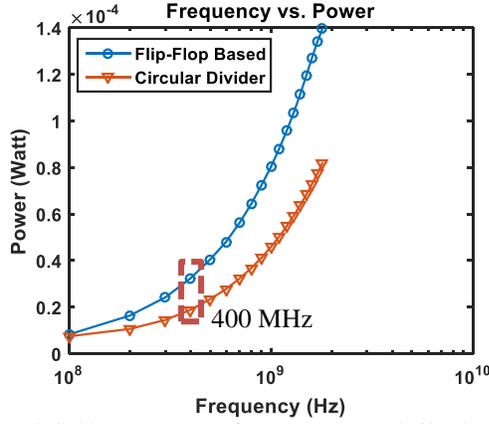

Figure 7: 2 dividers structure performance compared. Circular divider of [16] showed better power performance at 400MHz

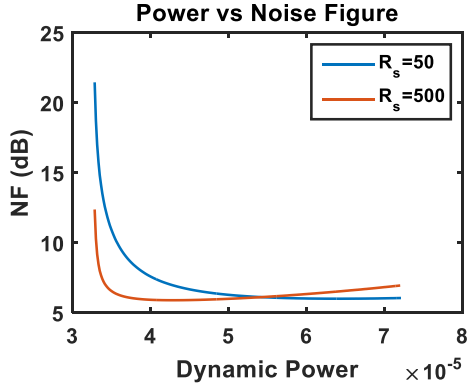

Figure 8: Dynamic power of 4-path mixer vs noise figure achieved under 2 different source impedance. This variation shows that to improve mixer-first's performance it is crucial to transform source impedance from 50 to larger values.

The power consumption of the circular divider along with the passive mixer is simulated with supply voltage of 1V and a load capacitance of 5pF. Figure 8 shows noise figure tradeoff with dynamic power of the mixer. The divider and MOS switches were simulated with a fixed supply voltage and the variation in dynamic power is obtained through variation in MOSFET size, as the gate capacitance of the switches act as extra capacitive load. From Figure 8, it can be seen that increasing dynamic power beyond a certain point does not improve noise figure. Simulation results show that the optimum point of operation corresponds to switch size of 10μm. Figure 8 also suggests larger source impedance improves noise performance. From [19], the noise figure can be simplified to Eq. (4) and that minimizing switch resistance $R_{sw}$ and maximizing source resistance $R_s$ would optimize noise performance. However, $R_{sw}$ is inversely proportional to transistor width, which varies with power consumption proportionally. From power budgeting prospective, increasing $R_s$ should be prioritize over reducing $R_{sw}$.

$$NF = \frac{\pi^2}{4} \frac{1+\frac{R_{sw}}{R_s}}{1-\frac{R_{sw}}{R_s}} \qquad (4)$$

Since power consumption in a digital circuit is proportional to its frequency and the band of operation of this receiver is 400MHz, relatively low compared to the 2.4GHz band, a digital mixer is ideal for this implementation. Although the NLO had

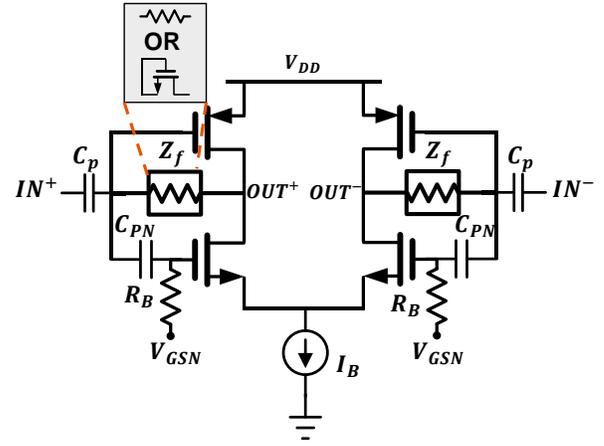

Figure 9: Differential Baseband LNA employing resistive feedback or LNA employing off-MOS feedback if bandwidth operation less than 1MHz. Outputs were loaded with 150fF capacitor to account for loading by latter stages of systems

4 25% duty cycle, 2 of the phases (0 and 180) were chosen to use for the down-conversion since OOK modulation does not utilize quadrature-phase to transmit separate message. In addition, 25% duty cycle clock was chosen, since it provides better selectivity compared to two phases with 50% duty cycle.

### III. RECEIVER ARCHITECTURE AND SIMULATED PERFORMANCE

#### A. Complete Receiver

With the system component's power and operable frequency range closely examined, it can be easily concluded that in order to avoid overhead bandwidth, optimal design should be implemented with passive matching instead of active matching, and baseband amplification should be strictly control to only support up to the bandwidth of the allocated transmission band.

For this front-end implementation, a mixer first architecture was adopted. The circular divider presented in [20] was used to construct the front-end. As mentioned, 2 path passive mixer was used instead of 4. Since passive mixer splits signal differentially, the LNA is designed into differential form as shown in Figure 9, while using the same biasing condition as depicted in section II A. For LNA operation bandwidth lower than 1MHz, off-transistors were used instead of feedback resistor as shown in Figure 9. The LNA outputs were loaded with 150fF capacitor to account for loading by latter stages of systems.

Operating under OOK scheme with DDR transmission signal input, the receiver is designed to be zero-IF, hence the LO frequency is equal to the carrier frequency. The receiver front-end is shown in the red dashed box in Figure 10. The system utilizes a 50 ohm off-chip resistor in the front end and uses off-chip matching network to perform step-up impedance transformation seen by the mixer, improving noise performance according to Eq. (4). While using high resistance at the mixer switch's input in conjunction with step-down matching is theoretically equivalent, large on-chip resistor are less desirable. The inductor $L_m$ used was 180nH and $C_m$ was 2pF, simulated along with the front-end but envisioned to be implemented off-chip.

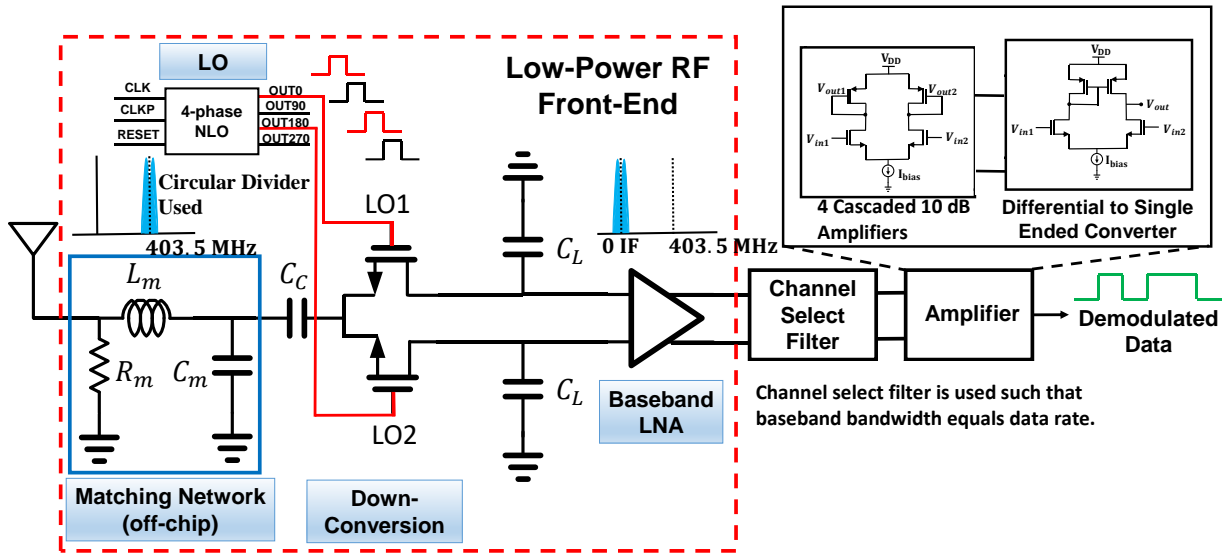

Figure 10: Low Power Front-end System. Components in the red dashed box is the simulation scope of this paper. 50 ohm $R_m$ matches to antenna and the step-up transformation seen from mixer's input improves noise performance and avoids large on-chip resistor. 2 path mixer was used as OOK doesn't utilize signal's phase to transmit information. Channel select filter and 4 cascaded 10 dB gain amplifier were used to test the functionality.

The bandwidth of LNA is set to the spectrical occupancy of a desired data rate. For MedRadio compliant, ideally the LNA would be limited to 300 kHz bandwidth. High data rate mode was also simulated as motivated in section I.

### B. Simulation Results

The receiver front-end of Figure 10 was simulated with TSMC 65nm in SPECTRE. OOK signals inputs were also fed as input to the front-end. To test the functionality, channel select filter and 4 cascaded 10-dB gain amplifiers were used, providing enough gain to verify the demodulated data, but are not in the simulation scope of this paper. Power consumption and sensitivity were estimated with PSS and PNOISE simulation tools. 8bit-10bit coding was used such that majority of baseband information's energy is located at the Nyquist frequency. Noise figure was simulated and measured at this frequency. The theoretical sensitivity then is estimated through equation (5)

$$-174 + 10Log_{10}(BW) + SNR_o + NF \qquad (5)$$

The envisioned channel selection filter would pass only the desirable bandwidth hence $BW$ was taken as the bandwidth of the channel where the data occupies, making the two equivalent. The $SNR_o$ term represents the signal to noise power ratio for a given BER under a modulation scheme. While modulation BER performance is given in $E_b/N_0$ form, it can be converted to SNR power ratio as shown in (6)[21].

$$\frac{S}{N} = \frac{R \times E_b}{BW_{data} \times N_0} \qquad (6)$$

$E_b$ is energy per bit, $N_0$ is noise spectral density and $R$ is data rate. For the front-end aiming to utilize OOK, the data rate is the same as data occupied baseband bandwidth with DDR. As shown in [22] and with Eq. 6, OOK with BER of $10^{-3}$ requires about 13dB SNR. Figure 11 shows the possible deign point simulated for the mixer-first receiver front-end. The theoretical sensitivities were estimated at different system bandwidth operation for optimal bandwidth utilization.

For current standard compliant receiver, the front-end is also adopted with LNA operated with 1MHz bandwidth for optimal sensitivity. Motivated by the fact that in Figure 12, system

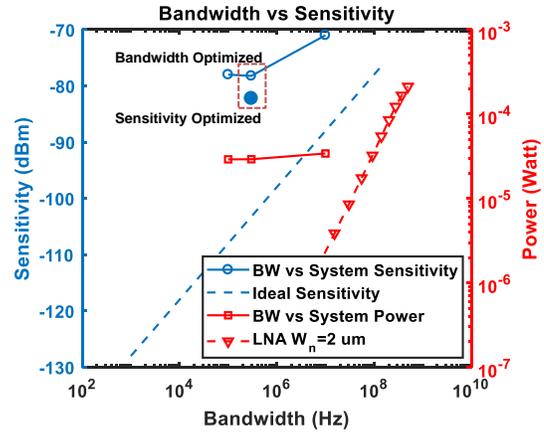

Figure 11: Bandwidth vs Sensitivity and Power required. Ideal sensitivity tendency on bandwidth and LNA power variation is shown for comparison. BW and sensitivity optimal design are shown separately as annotated. Standard compliant performance shown in red dashed box

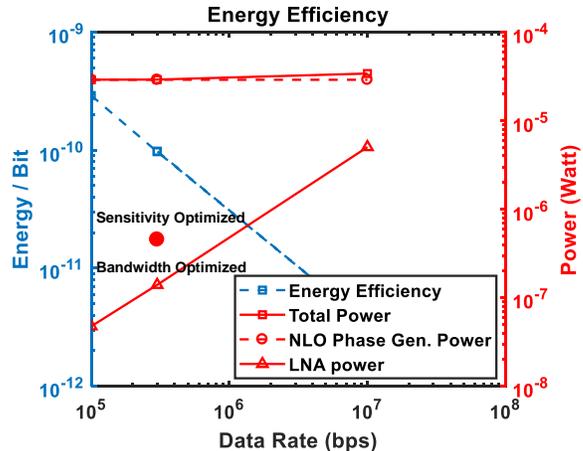

Figure 12. Bandwidth vs Energy per Bits. Data rate reduction does not improve energy efficient infinitely as power is limited by LO synthesis. Phase gen. power exhibits no data rate dependency yet total power is limited.

TABLE I : FRONT END PERFORMANCE COMPARISON

| Symbol | [20] | [21] | | [14] | [19] | This work (Simulation) | |
|---|---|---|---|---|---|---|---|
| Receiver Purpose | MICS Transceiver | WuRx | | MICS Front-end | MICS Front-end | MedRadio Front-end | |
| Architecture | RF LNA | Mixer-First | | RF LNA | RF LNA | Mixer-First | |
| Modulation | OOK | OOK | | - | FSK | OOK | |
| Supply Voltage | 1.8 | 0.75 V | | 1V | 1.8 V | 1 V | |
| Frequency | 400MHz | 2.45 GHz | | 401-457MHz | 402-405MHz | 402-405MHz | |
| Power | 3.4 mW | 50 µW | | 370µW | 1.3mW | 29.2µW | 34 µW |
| BW/Data Rate | 2Mbps | 250kbps | 650kbps | 1Mbps | 200kbps | 300kbps* | 10Mbps* |
| Energy Efficiency | 1.71 nJ/b | 200pJ/b | 77pJ/b | 370pJ/b | 6.5nJ/b | 97pJ/b | 3.4pJ/b |
| Sensitivity (dBm) $10^{-3}$ BER | -80.2 | -88 | -71 | -96(if were to use OOK) | -96.8 | -83# | -70# |
| Technology | 0.18µm | 65nm | | 0.18µm | 0.18 µm | 65 nm | |

*OOK DDR signaling potentially supporting up to 300kbps data rate under 300 kHz bandwidth limitation, or 10Mbps under 10MHz bandwidth, here is used to estimate the sensitivity and energy efficient, #No implementation loss included and should be added into design consideration

power ceases to reduce at 300kbps, but noise figure from LNA continues to degrade. This translates to -70dBm sensitivity as shown in the red dashed box in Figure 11, consuming about 34 µW. The sensitivity for optimal bandwidth setting, LNA having 300kHz bandwidth, were also shown in Figure 11. In Table 1, the 10Mbps results are compared with other literature reported in [23], [24], [25]. The energy efficiencies were estimated under the desired data rate.

In Figure 12, one can observe that although decreasing bandwidth at circuit level reduces power, at system level, it does not translate to better energy efficiency. This is shown with the power break-down of the system simultaneously. When trying to reduce power through bandwidth adjustment, low bandwidth operation is eventually limited by the LO phase generation circuit's power hence the energy/bit does not improve after certain operation data-rate.

## IV. CONCLUSION

Implantable healthcare monitoring calls for ultra-low active power (<50µW) high data-rate, energy-scalable, highly energy-efficient (pJ/bit) radios. Investigation of power performance trade-off of each front-end component in a conventional radio reveals 50Ω active matching and RF gain is prohibitive for 50µW power-budget. Receiver front-end designed with N-path mixer and baseband LNA in TSMC 65nm technology achieved sensitivity -83 dBm sensitivity and <100 pJ/b for MedRadio compliant data rate and -70 dBm sensitivity with <5pJ/b at 10Mbps. Since LO synthesis power dominated the total power, future work would focus on energy/performance scalable LO and high-impedance interfaces.

## V. ACKNOWLEDGEMENT

The work is supported by Semiconductor Research Corporation (SRC) Grant No. 2720.001 and National Science Foundation (NSF) CRII Award, CNS Grant No. 1657455.

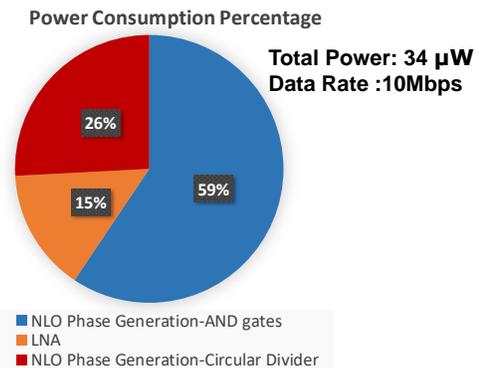

Figure 13. Power distribution amongst key front-end components. LO synthesis components occupy the majority of the power.

## VI. REFERENCES

[1] C. M. Lopez et al., "An Implantable 455-Active-Electrode 52-Channel CMOS Neural Probe," IEEE JSSC 2014
[2] "PillCam Capsule Endoscopy - Given Imaging." [Online]. Available: http://www.givenimaging.com/en-int/Innovative-Solutions/Capsule-Endoscopy/Pages/default.aspx. [Accessed: 31-Jul-2017].
[3] D. Panescu, "Emerging Technologies [wireless communication systems for implantable medical devices]," IEEE Eng. Med. Biol. Mag 2008.
[4] P. D. Bradley, "An ultra low power, high performance Medical Implant Communication System (MICS) transceiver for implantable devices," in BioCAS 2006.
[5] S. Sen, "Invited: Context-aware energy-efficient communication for IoT sensor nodes," in DAC 2016
[6] S. Sen, et al., "TRIFECTA: Security, Energy-Efficiency, and Communication Capacity Comparison for Wireless IoT Devices," IEEE Internet Comput., In Press.
[7] S. Sen, "SocialHBC: Social Networking and Secure Authentication using Interference-Robust Human Body Communication," in IEEE ISLPED 2016.
[8] S. Maity et al., "Adaptive interference rejection in Human Body Communication using variable duty cycle integrating DDR receiver," in DATE), 2017
[9] S. Mait et al., "Secure Human-Internet using dynamic Human Body Communication," in 2017 IEEE/ACM ISLPED 2017
[10] "Medical Device Radiocommunications Service (MedRadio)," Federal Communications Commission, 09-Dec-2011. [Online]. Available: https://www.fcc.gov/wireless/bureau-divisions/broadband-division/medical-device-radiocommunications-service-medradio.
[11] H. T. Friis, "A Note on a Simple Transmission Formula," Proc. IRE, vol. 34, no. 5, pp. 254–256, May 1946.
[12] A. J. Johansson, "Performance of a radio link between a base station and a medical implant utilising the MICS standard," in The 26th Annual International Conference of the IEEE Engineering in Medicine and Biology Society, 2004
[13] H. K. Cha, et al., "A CMOS MedRadio Receiver RF Front-End With a Complementary Current-Reuse LNA," TMTT 2011
[14] T. Taris, et al. "A 60µW LNA for 2.4 GHz wireless sensors network applications," in RFIC 2011
[15] M. Parvizi, et al., "A 0.4V ultra low-power UWB CMOS LNA employing noise cancellation," ISCAS2013
[16] M. Parvizi, et al., "Short Channel Output Conductance Enhancement Through Forward Body Biasing to Realize a 0.5 V 250 µW 0.6-4.2 GHz Current-Reuse CMOS LNA," IEEE JSSC 2016
[17] M. Parvizi, et al., "An Ultra-Low-Power Wideband Inductorless CMOS LNA With Tunable Active Shunt-Feedback," TMTT 2016.
[18] C. Choi, et al., "A 370µW CMOS MedRadio Receiver Front-End With Inverter-Based Complementary Switching Mixer," MWCL 2016
[19] C. Salazar, et al., "A 2.4 GHz Interferer-Resilient Wake-Up Receiver Using A Dual-IF Multi-Stage N-Path Architecture," IEEE JSSC 2016
[20] A. Ba et al., "A 1.3 nJ/b IEEE 802.11ah Fully-Digital Polar Transmitter for IoT Applications," IEEE JSSC 2016
[21] D. Terlep, "Receiver Sensitivity Equation for Spread Spectrum Systems." Maxim Integrated Products, Inc, 28-Jun-2002.
[22] Q. Tang et al., "BER performance analysis of an on-off keying based minimum energy coding for energy constrained wireless sensor applications," ICC 2005
[23] H. Cruz, et al. "A 1.3 mW low-IF, current-reuse, and current-bleeding RF front-end for the MICS band with sensitivity of -97 dbm," IEEE TCASI 2015
[24] Y.-H. Liu, et al. "A low-power asymmetrical MICS wireless interface and transceiver design for medical imaging," in BioCAS 2006.
[25] C. Bryant et al., "A 2.45GHz, 50uW wake-up receiver front-end with -88dBm sensitivity and 250kbps data rate," in ESSCIRC 2014